\documentclass[aps,preprintnumbers,twocolumn]{revtex4}
\usepackage{amsfonts,amssymb,amsmath}            
\usepackage[]{graphics,graphicx,epsfig}            
\usepackage{amsthm}                             
\def\identity{\leavevmode\hbox{\small1\kern-3.8pt\normalsize1}}

\newcommand{\ket}[1]{\left | #1 \right\rangle}

\newcommand{\half}{\mbox{$\textstyle \frac{1}{2}$}}

\begin{document}

\title{State Transfer and Spin Measurement}
\date{\today}

\author{Alastair \surname{Kay}}
\affiliation{Centre for Quantum Computation,
             DAMTP,
             Centre for Mathematical Sciences,
             University of Cambridge,
             Wilberforce Road,
             Cambridge CB3 0WA, UK}
\begin{abstract}
We present a Hamiltonian that can be used for amplifying the signal from a quantum state, enabling the measurement of a macroscopic observable to determine the state of a single spin. We prove a general mapping between this Hamiltonian and an exchange Hamiltonian for arbitrary coupling strengths and local magnetic fields. This facilitates the use of existing schemes for perfect state transfer to give perfect amplification. We further prove a link between the evolution of this fixed Hamiltonian and classical Cellular Automata, thereby unifying previous approaches to this amplification task.

Finally, we show how to use the new Hamiltonian for perfect state transfer in the, to date, unique scenario where total spin is not conserved during the evolution, and demonstrate that this yields a significantly different response in the presence of decoherence.
\end{abstract}

\maketitle

One of the many challenging tasks in realising technology for quantum information processing is measuring the output of a protocol. Typically, the result is expected to be stored on single spins, when even detecting the presence of a single spin is an experimental challenge, let alone measuring its internal state. Recently, there have been proposals for amplifying a quantum state so that it is converted into a macroscopic property which can be measured \cite{lee:05,cappellaro:05,mosca:06}. Since copying the state many times is impossible, the aim of these protocols is to perform the conversion
\begin{equation}
(\alpha\ket{0}+\beta\ket{1})\ket{0}^{\otimes(N-1)}\rightarrow\alpha\ket{0}^{\otimes N}+\beta\ket{1}^{\otimes N}.	\label{eqn:target}
\end{equation}
If the states $\ket{0}$ and $\ket{1}$ are stored on magnetic sub-levels of the spins, this yields a macroscopic magnetisation which can be measured to determine if the original state was $\ket{0}$ or $\ket{1}$. The original proposal focused on using a fixed Hamiltonian that achieved some macroscopic change, although did not manage the desired transformation with unit fidelity. Subsequently, the idea of using techniques from classical Cellular Automata (CA) was proposed \cite{mosca:06}, which realised a speed-up by using a cubic organisation of spins instead of a linear geometry.

In this paper, we provide a unification of these techniques in a one-dimensional system. Making use of existing work on perfect state transfer in spin chains, primarily from \cite{Bos03,Christandl,Christandl:2004a,Kay:2004c}, we can rephrase a large set of results in the context of this system. This leads us to develop a fixed Hamiltonian which not only achieves the evolution in Eqn.~(\ref{eqn:target}), but accurately reconstructs the results of the cellular automaton for all possible initial states of the system. This new Hamiltonian can, itself, be used for perfect state transfer without any external interaction, and is the first example that does not preserve the total spin during the process. As such, its behaviour can be expected to be significantly different when, for example, noise is present.

In the one-dimensional case, the way the CA approach worked was to apply a series of commands to the chain of spins. These commands were capable of detecting a local sequence of either $\ket{1x0}$ or $\ket{0x1}$ and converting them into $\ket{1{\bar{x}}0}$ and $\ket{0{\bar{x}}1}$ respectively. Ensuring that $x=0$ guarantees amplification (an increase in the number of 1's, providing there was already at least one 1 present). This was achieved by alternating the application of these pulses to even and odd qubits on the chain. There is a corresponding Hamiltonian which can achieve the same result,
$$
K_n=\half(X_n-Z_{n-1}X_nZ_{n+1}).
$$
The CA commands that are applied at either end of the chain are slightly different. Since we never want to flip the first qubit, there is no first term. At the end of the chain, we want to convert from $\ket{10}$ to $\ket{11}$, hence we set
$$
K_N=\half(\identity-Z)_{N-1}X_N.
$$
The CA proceeds by alternately applying $\sum_nK_{2n}$ and $\sum_nK_{2n+1}$. However, we wish to create a fixed Hamiltonian that does not require this alternation of terms. As noted in \cite{lee:05}, the Hamiltonian
$$
H=\sum_{n=2}^{N}J_{n-1}K_n
$$
keeps the state $\ket{100\ldots 0}$ in a subspace which we describe as
\begin{equation}
\ket{\tilde n}:=\ket{\overbrace{1\ldots 1}^n\overbrace{0\ldots 0}^{N-n}}. \label{eqn:redefine}
\end{equation}
Calculating the action of the Hamiltonian, $H\ket{\tilde n}=J_{n-1}\ket{\widetilde{n-1}}+J_n\ket{\widetilde{n+1}}$, we observe that this is identical to the action of an exchange Hamiltonian
$$
H_\text{ex}=\half\sum_nJ_n(X_nX_{n+1}+Y_nY_{n+1})
$$
on a single excitation located on site $n$, $H_\text{ex}\ket{n}=J_{n-1}\ket{n-1}+J_n\ket{n+1}$. Hence, if we were to set $J_n=1$, we would immediately recover the results of \cite{Bos03,Kay:2004c} for state transfer using a uniformly coupled chain ($\ket{1}\rightarrow\ket{N}$), but in the case of signal amplification. In particular, we recover the same calculations as \cite{lee:05}, and the fact that the goal of Eqn.~(\ref{eqn:target}) can never be perfectly realised in chains of more than 3 spins with uniform couplings. We also discover how to perfectly realise the process $\ket{\tilde{1}}\rightarrow\ket{\tilde{N}}$, by engineering the couplings of terms to $J_n=\sqrt{n(N-n)}$, while leaving the state $\ket{0}$ unchanged \cite{Christandl,Kay:2004c}.

Many other results from the study of state transfer can readily be applied to obtain perfect, or near-perfect, amplification under different constraints, depending on what can physically be achieved. For example, a local magnetic field \cite{shi:2004} could be used to enhance the amplification on the uniformly coupled Hamiltonian \cite{lee:05}, where we replace the local magnetic fields $\sum B_nZ_n$ of the exchange model with either $\sum B_nZ_nZ_{n+1}$ terms \footnote{This is the strict equivalence that holds for all excitation subspaces.} or local magnetic fields $\sum B_n'Z_n$ such that $B_n'=B_{n-1}-B_n$. Alternatively, we could tune the couplings $J_n$ and magnetic fields $B_n$ to yield different spectra \cite{transfer_comment, Kay:2005e}, which enable control of a range of useful properties such as the robustness against a variety of errors. Naturally, schemes that require single spin measurement to herald the correct evolution, such as \cite{Bos04,Bose:2005a}, should not be used.

We have demonstrated a mapping between the Hamiltonians $H$ and $H_\text{ex}$ in the $0^{th}$ and $1^{st}$ excitation subspaces. The equivalence is not limited just to these subspaces, however. For example, the second excitation subspace of $H_\text{ex}$ is denoted by $\ket{n,m}$, which describes excitations at the sites $n$ and $m$. We can easily demonstrate that the action of $H$ on the state $$\ket{{\tilde{n}},{\tilde{m}}}:=\ket{{\tilde{n}}\oplus{\tilde{m}}}$$ is identical (for example, $\ket{{\tilde{3}},{\tilde{5}}}=\ket{11100\oplus 11111}=\ket{00011}$). The generalisation to higher excitation subspaces is straightforward, constituting bitwise addition modulo 2 of the effective single excitations. Formally, we can prove the equivalence of the two Hamiltonians by demonstrating a transformation between them \footnote{We thank Michael Nielsen for suggesting this method.}. We repeatedly apply controlled-NOT gates $C_n^{n-1}$ to $H_\text{ex}$ with control qubit $n$ and target qubit $n-1$ starting from $n=N$ and finishing with $n=2$,
$$
C_2^1C_3^2\ldots C_N^{N-1} H_\text{ex} C_N^{N-1}\ldots C_2^1 = H.
$$
The equality with $H$ follows from the standard propagation properties of the Pauli matrices through the controlled-NOT gate \cite{nielsen}, where $Z$ propagates from target to control and $X$ from control to target. Hence, the $X_nX_{n+1}$ terms become $X_{n+1}$ and the $Y_nY_{n+1}$ terms become $-Z_nX_{n+1}Z_{n+2}$, except for the final term, which transforms into $-Z_{N-1}X_N$, thereby recovering $H$.  The same transformation can also be used to show how local magnetic fields transform, and subsequently allows us to describe the subspace structure of $H$,
$$
\left[H,\sum_{n=1}^{N-1}Z_nZ_{n+1}+Z_N\right]=0.
$$

As a consequence of the equivalence of $H$ and $H_\text{ex}$, not only can we use results for the single excitation subspace of state transfer chains, but for all excitation subspaces \cite{Christandl:2004a}. In particular, if we use the engineered couplings $J_n=\sqrt{n(N-n)}$, we find that for any classical initial state of the chain (not just $\ket{0}^{\otimes N}$ and $\ket{1}\ket{0}^{\otimes (N-1)}$), we get a classical output, because perfect state transfer occurs in all excitation subspaces. This output is precisely that given by the cellular automaton. The action of a CA command is
$$
\ket{x,y,z}\rightarrow\ket{x,x\oplus y\oplus z,z}
$$
on every second qubit. By construction, the action of our Hamiltonian on the effective single excitation subspace corresponds to the CA. Therefore, and as a consequence of the facts that bitwise addition operation is commutative, and our basis states are correctly described by bitwise addition, the output must be the same as the CA for the whole space of states.

The Hamiltonian $H$ is, up to the local terms $\half\sum_{n=2}^NX_n$, the cluster state Hamiltonian i.e.~the Hamiltonian that has the cluster state as its ground state. It has been proven \cite{nielsen:05}, that 3-body terms are a necessity for any such Hamiltonian. This must be true, not only for the ground state, but also for the excited states because the excited states can be turned into ground states by local $Z$ operations. Consequently, we can conclude that any Hamiltonian which is to give the same evolution as $H$ must consist of at least 3-body terms. This is true whatever the coupling strengths $J_n$ because $[K_n-\half X_n,K_m-\half X_m]=0$ and hence the coupling strengths only determine the spectrum of the cluster state Hamiltonian, not the eigenstates themselves.

\begin{figure}[!t]
\begin{center}
\includegraphics[width=0.5\textwidth]{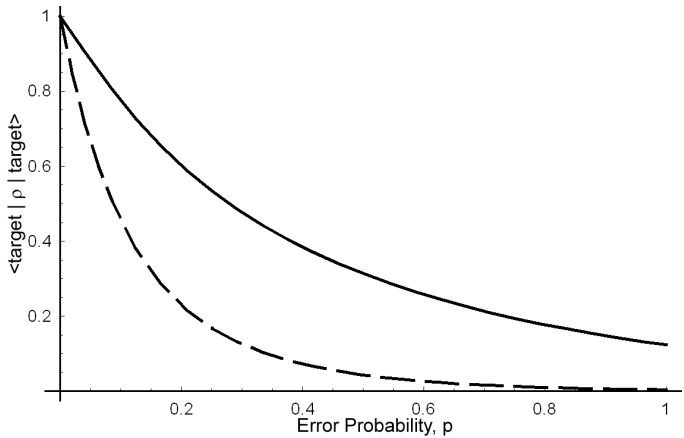}
\end{center}
\caption{Comparison of state transfer across 6 spins using $H$ (solid line) and $H_\text{ex}$ (dashed line) in the presence of dephasing noise. We have assumed that a single phase flip occurs with the probability $p$ at each of 25 time steps on a random qubit.
The initial state is a single excitation, and the measured fidelity is the probability that an excitation is found on the expected output qubit.} \label{fig:decoherence}
\end{figure}
With the definition of higher excitations as stated, we can readily see that our Hamiltonian also performs perfect state transfer. A single excitation at site $n\neq 1$ is in the state $\ket{\widetilde{n-1},\tilde{n}}$, and transfers to $\ket{\widetilde{N+1-n},\widetilde{N+2-n}}$. The effect of fermion exchange \cite{bose:2004a,Christandl:2004a} now manifests itself as a $Z$ gate on the state. Furthermore, we can choose to transfer several states at once, and while perfect transfer still occurs, there is no controlled-phase gate between the exchanging states. While potentially useful for some purposes \cite{Jaksch:2004a,Kay:2004c}, it is undesirable in the case of state transfer, so this Hamiltonian provides an alternative to encoding the qubits in pairs of spins $\ket{0_L}=\ket{01}$ and $\ket{1_L}=\ket{10}$ on the same chain, as is required for all previous chains. This unique behaviour, which can be viewed as a consequence of the fact that the Hamiltonian is not spin preserving, can be expected of affect other properties of the chain. In particular, one might expect it to have significantly different resistance to decoherence than previous chains. In Fig.~\ref{fig:decoherence}, we have examined dephasing noise applied during state transfer using the two different Hamiltonians and observe that the new Hamiltonian demonstrates a significant enhancement in robustness.

\begin{figure}[!t]
\begin{center}
\includegraphics[width=0.25\textwidth]{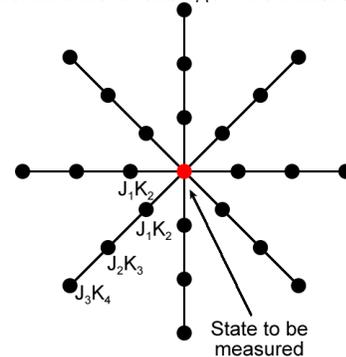}
\end{center}
\caption{Star-shaped geometry which consists of the spin the be measured in the centre, and a set of one-dimensional chains radiating outwards. This works because the only terms that are applied to the central spin are $Z$, which commute. Note that the coupling strengths do not need to be reduced, as in \cite{Kay:2005b}.} \label{fig:star}
\end{figure}

With the exception of the star-shaped construction depicted in Fig.~\ref{fig:star}, these results do not extend to higher dimensional geometries. There are two main reasons for this. Firstly, there are no known geometries other than a simple chain that perfectly transfer states in all excitation subspaces at the same time \cite{Kay:2005b}. Secondly, even the chain splitting techniques demonstrated in \cite{Kay:2005b}, and rediscovered in \cite{yang:05}, do not work with the new Hamiltonian if we restrict to 3-body terms. This splitting technique worked by observing that if two spins are each coupled to a single spin with strength $J/\sqrt{2}$, then the state $\ket{01}+\ket{10}$ across this pair can be treated by replacing the pair of qubits with a single qubit, coupled with strength $J$, and a single excitation. Using this technique again, we still get a superposition of $\ket{01}+\ket{10}$, and not the $\ket{11}$ which would be required for amplification. If we were to use the geometry illustrated in Fig.~\ref{fig:star}, then the signal gets enhanced by a factor $R$, where there are $R$ spikes on the star. The different spikes do not compete with each other because the Hamiltonians $H$ only interact on the qubit where the state is initially stored, and on this qubit, they commute because all terms are $Z$ and $\identity$. Alternatively, we could say that to get a particular signal size, we require $N$ qubits, and hence the time required for the protocol is reduced to $N/R$, which is unable to match the cubic geometries possible in CA systems, which only require a running time of $\sqrt[3]{N}$ \cite{mosca:06}. 

In summary, we have presented a map between an exchange Hamiltonian and a cluster-like Hamiltonian which enables previous ideas on perfect state transfer to be directly applied to the problem of signal amplification. It has also yielded a new method for state transfer, which is not restricted to being spin preserving, and avoids the problems of fermionic exchange when multiple states are transferred. Intriguingly, the presented Hamiltonian implements a discrete CA in a continuous-time system.

The present work raises several interesting questions. For example, is it possible to find Hamiltonians that correctly simulate Quantum Cellular Automata which issue commands such as ``apply the operation $U$ if your neighbours are different'', perhaps by using a new Hamiltonian of the form
$$
H_\text{QCA}=\sum_nJ_{n-1}(H_n-Z_{n-1}H_nZ_{n+1}),
$$
where the coupling strengths $J_n$ and local Hamiltonians $H_n$ would need to be determined. The existing work at least justifies that there will be a similar subspace restriction. On a more wide-reaching basis, we have succeeded in demonstrating that state transfer ideas need not be restricted to spin preserving Hamiltonians, as previously thought. Consequently, are there any other useful protocols to which we can apply similar ideas? One potential candidate is the optimal universal cloning machine. Previous attempts, when restricted to spin preserving Hamiltonians, have only succeeded in creating phase-covariant cloners \cite{spin_chain_clone}.

The author is supported by the UK EPSRC and Clare College, Cambridge. He would like to thank D. Burgarth for useful discussions.

\end{document}